\documentclass{article}

\usepackage{arxiv}

\usepackage{mathtools}

\usepackage[utf8]{inputenc} 
\usepackage[T1]{fontenc}    
\usepackage{hyperref}       
\usepackage{url}            
\usepackage{booktabs}       
\usepackage{amsfonts}       
\usepackage{microtype}      
\usepackage{graphicx}
\usepackage[super]{natbib}
\usepackage{nccmath}
\usepackage{listings}

\title{An introduction to group sequential methods: planning and multi-aspect optimization}

\author{ \href{https://orcid.org/0000-0003-4580-2712}{\includegraphics[scale=0.06]{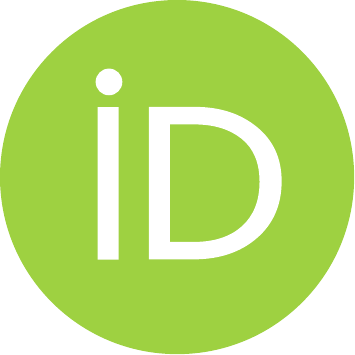}\hspace{1mm}Fraser I.~Lewis}
\\GSK\\
	\texttt{fraser.i.lewis@gsk.com} 
}

\renewcommand{\shorttitle}{Group Sequential Methods Tutorial}

\hypersetup{
pdftitle={An introduction to group sequential methods: planning and multi-aspect optimization},
pdfsubject={statistics},
pdfauthor={Fraser I.~Lewis},
pdfkeywords={Clinical Trial, Group Sequential, Tutorial},
}

\begin{document}
\maketitle

\begin{abstract}
A group sequential clinical trial design can be an attractive option when planning a pivotal trial as this approach has the ability to stop the trial early for success, whilst also being well accepted from a regulatory review perspective. Compared to a single stage design there are more moving parts to consider and optimise when planning a group sequential trial. This tutorial briefly outlines the group sequential methodology before detailing some of the key operating characteristics and how these can be estimated, optimised and ultimately presented to decision makers when aligning on a final study design.   
\end{abstract}


\section{Introduction}\label{sec1}
In many therapy areas pivotal clinical trials typically follow a classical single stage design, where patient data are accumulated throughout the trial, and only once the trial is complete are the data unblinded and the pre-planned statistical analyses executed. Group sequential clinical trial designs extend this single stage design to a multi-stage design, where at each stage an unblinded pre-planned analysis can be conducted on interim accumulations of the data. These interim analyses may permit early stopping of the study for efficacy or futility. Group sequential designs, therefore, can be effective risk mitigation strategies against either unexpectedly poor results (and so a stop for futility) or unexpectedly very positive results, e.g. where highly persuasive evidence is already available at an interim look at the data, thereby a stop for efficacy. Whilst group sequential designs are well established in the literature, practical experience in the planning and optimisation of such designs for a pivotal study is not widespread. This article comprises a case study-based tutorial on how to design and optimise a group sequential design from first principles without the use of specialist clinical trial software or libraries. Supplementary material is provided which uses R code for the various calculations required. 

We focus on a specific use case where the clinical research question is: what is the difference in treatment effect between a test regimen and standard of care, where each patient is either a responder or non-responder (a binary endpoint), and where study success is defined as providing sufficient evidence to reject a null hypothesis of inferiority of the test regimen to the standard of care. Each of the design and calculation steps later discussed equally apply to other situations, for example continuous endpoints and superiority testing, but the technical specifics will differ due to the different test statistics.  

Group sequential methods are well suited to pivotal trials and this is our implicit focus here. Given the resources required for a pivotal trial and the likely subsequent regulatory review, then the approach to risk management can arguably be rather different from earlier phase studies. We consider here a single interim analysis, but the methodology and calculations naturally extend to more than a single interim. The number of interim looks, particularly in a pivotal trial, requires careful consideration in terms of additional operational complexity and risks associated with repeatedly unblinding data during study conduct. 
 
In the following sections we first briefly outline some of the key literature on group sequential methods, along with some alternative approaches, before giving an overview of the key design features and decisions required. We then detail the operating characteristics of these designs and present some examples of the calculations required, demonstrating the roles of the different design parameters. In the results section we work through our case study step-by-step showing how to compute the various numerical results needed and demonstrate the multi-aspect nature of choosing an optimal group sequential design. We conclude with a short discussion. 

\subsection{Brief literature review on group sequential methods}
Group sequential methods in clinical trials have a long history, comprising an already large body of work in the 20 years prior to the now standard text in this field ``Group Sequential Methods: Applications to Clinical Trials'' by Christopher Jennison and Bruce Turnbull\citep{Jennison:2000aa}, which is itself now over 20 years old. Group sequential methods also feature in ICH guidance and in the FDA's ``Adaptive Designs for Clinical Trials of Drugs and Biologics: Guidance for Industry''  (November 2019).  Group sequential methods are sufficiently well tried and tested that they are no longer considered a novel or complex design methodology, and so would not typically warrant prior scientific consultation with regulatory authorities. This is generally in contrast to adaptive designs, where major data-driven design changes are possible during study execution (including interim unblinded looks at the data), which introduces additional risks and complexities into a pivotal trial design (e.g. see Bauer\citep{Bauer:2016aa} and Wassmer\citep{Wassmer:aa}). In a standard group sequential design there are no data-driven adaptations, rather all decisions regarding stop/continue based on interim unblinded looks at the data are based on pre-specified criteria, this ensures strict control of the type I error rate. 

Compared to single stage designs group sequential designs require a larger sample size to achieve the same power all else being equal, as there is an inevitable statistical cost for repeated analyses. Depending on the design choices made in the group sequential design the increase in sample size over a single stage design can be modest but is also a trade off with other operating characteristics. The comparative efficiency of different group sequential designs verses single stage designs is extensively covered in existing literature, including in Jennison and Turnbull \citep{Jennison:2000aa}. Here we focus on how to choose group sequential designs to achieve given desirable operating characteristics.

There are non-adaptive alternatives to group sequential methods for pivotal studies which include interim unblinded looks at the data, with arguably the main competitor being the ``Promising Zone'' approach by Mehta\citep{Mehta:2011aa,Mehta:2016aa}. The promising zone design is based on a set of rules which allows the sample size in the study to be increased by a certain margin based on the results of an unblinded interim look at the data. This adaptation in sample size is not something which is typically possible in group sequential design (but see Cui\citep{Cui:1999aa} for an exception). As the name suggests, if the unblinded interim results fall within a particular zone - the promising zone - then the sample size can be increased by a certain amount without any detrimental impact on the type I error. This ease of use makes such a design attractive for pivotal studies. A number of articles have compared group sequential and promising zone designs\citep{Jennison:2015aa,Edwards:2020aa}, and whether sample size re-estimation at interim is any more efficient in terms of expected sample size and power than a group sequential design is arguably an open question and case specific.

\section{Methods}
In this section we borrow extensively from Jennison\citep{Jennison:2000aa} and focus on the specific use case of a randomised controlled trial comprising of two treatment arms, with a binary primary endpoint and an objective of demonstrating non-inferiority. Each of the following subsections briefly details a part of the design or an operating characteristic of interest. In Section \ref{casestudy} we show numerical examples of the actual calculations involved, with implementation details and full R code deferred to the supplementary information, and illustrate the choices required in trading-off the different operating characteristics of the designs to achieve an overall optimal design. 

\subsection{Statement of hypothesis test}
As with any single stage study a primary hypothesis test is needed, the difference in a group sequential design is that this hypothesis is tested multiple times as we take more than one look at the data. It is immediately obvious, therefore, that control of the type I error rate is an essential component of a group sequential design.  For our case study the null and alternative hypotheses are defined as: 
\begin{fleqn}
\begin{align}
&&H_ 0&: p_t - p_c -\delta\le{0}\label{hyp1}\\
&\text{and}&\ H_ 1&: p_t - p_c -\delta>0;\label{hyp2}
\end{align}
\end{fleqn} 
where $p_t$ and $p_c$ are the proportion of responders under a test regimen and standard of care respectively, and $\delta$ is the non-inferiority margin. An efficacious treatment has a higher proportion of responders and so support for the alternative hypothesis increases as $p_t-p_c$ increases. 

\subsection{Test statistic definition}
There are a number of different test statistics which could be used for the hypothesis test in eqns (\ref{hyp1})-(\ref{hyp2}). We shall use the simplest; a standard Wald test statistic for comparing two proportions with inclusion of a non-inferiority margin. Alternatives could be score-based tests (e.g. see \citep{Miettinen:1985aa,Newcombe:1998aa}) whose inclusion into a group sequential design follows the same steps as the Wald test, but is rather more algebraically complex. The standard Wald test for comparing two proportions including a margin is given below: 
\begin{fleqn}
\begin{align}
&&Z_{i}&=(p_t-p_c-\delta)\sqrt{\left(\frac{p_t(1-p_t)}{n_{i}}+\frac{p_c(1-p_c)}{n_{i}}\right)^{-1}};\label{eqn1}\\[2pt]
&&Z_{i}&=\theta \sqrt{I_{i}}; \label{eqn2}\\[5pt]
&\text{and}&Z_ {i} &\sim N(\theta \sqrt{I_{i}},1);
\end{align}
\end{fleqn}
where $n_{i}$ is the per treatment sample size with $i$ indexing the look at the data. In a standard single stage design we have $n_i=N$, the overall sample size. Note that the Wald test statistic in (\ref{eqn1}) is written as a product of the ``natural estimator'', $\theta=p_t-p_c-\delta$, as features directly in the statement of the null hypothesis, and information, $I_{i}$, where the latter is defined as the reciprocal of the variance of the natural estimator (following Jennison\citep{Jennison:2000aa}). In group sequential methods test statistics are typically parameterised in terms of information as opposed to variance or standard error. This terminology also lends itself to accumulating data because as a trial accumulates more data then this is equivalent to gaining more information, as information is an increasing function of $n_i$ (as in eqn \ref{eqn1}). 

\subsection{General formulation of joint distributions}
In a single stage trial hypothesis testing is focused at a single fixed point -- the overall accumulation of data -- where we need to know the probability distribution of the test statistic $Z$ at sample size $N$. In a group sequential design we need to know how the random variable $Z_i$ behaves as \emph{data is accumulated}. This is because with multiple looks at the data, then multiple $Z_i$ are involved in any calculations in regard to target operating characteristics, e.g. type I error and power. Hence, what is needed is the joint distribution of the $Z_i$s for any set of indexes $i$. We formally define this joint distribution at the end of this section. Simulations of how $Z$ behaves as data are accumulated are presented in Section \ref{casestudy}.

\subsubsection{Definition of joint distribution}
The joint distribution of $Z_i$s across multiple looks at the data is the key analytical result needed to progress with group sequential methods. This is directly analogous to requiring the sampling distribution of the chosen test statistics, under null and alternative hypotheses, in a single stage design. The difference here is that we have multivariate sampling distributions because we have multiple dependent test statistics, one for each unblinded look at the data. This multivariate sampling distribution is needed under each of the null and alternative hypotheses. The joint distribution formulation presented next is due to Jennison\citep{Jennison:2000aa} p.49.

For a group sequential design with $K$ total looks at the data we have a vector $(Z_1,\dots,Z_K)$ of test statistics where 
 \begin{fleqn}
\begin{align}
&& (i)&\quad(Z_1,\dots,Z_K) \hspace{0.2cm}\text{is multivariate normal,}\label{eqj0} \\
&&(ii)&\quad E(Z_k)=\theta \sqrt{I_k}, \hspace{0.2cm} k=1,\dots,K,\label{eqj1}\\
&\text{and}&(iii)&\quad Cov(Z_{k_1},Z_{k_2} )= \sqrt{I_{k_1}/I_{k_2}}, \quad 1\le k_1 \le k_2 \le K.\label{eqj2}
\end{align}
\end{fleqn}
This is a general result with the same level of applicability as maximum likelihood estimation itself (see p.70 Jennison\citep{Jennison:2000aa}), however, these are of course asymptotic results, and this distributional approximation may be much less robust very early on in a trial where information levels are low. 

Our current use case utilizes a Wald test but the generality of the above distributional result means that the use of more sophisticated scored-based tests, e.g. Miettinen and Nurminen test for risk difference\citep{Miettinen:1985aa} also fits into this formulation, and shows the general versatility of the group sequential methodology.   
 
To summarise so far, we have: i) stated the hypothesis test; ii) the test statistics to be used; and iii) the probability distribution of the test statistics. We now define the operating characteristics of a group sequential design with a single interim look. The operating characteristics of a study design are probabilistic events whose occurence we wish to control at suitably high or low levels in order to give a well controlled and successful study. These events are defined on the joint distribution in eqns (\ref{eqj0})-(\ref{eqj2}). We consider only a single interim look but the results naturally extend to multiple interim looks.

\subsection{Design operating characteristics}

\subsubsection{Error rates and critical values}
In any study design the type I and type II error rates are obviously of crucial importance, with type I error control being especially critical in a pivotal study. These errors rates are determined by critical values, or rather, when designing a study critical values are sought such that error rates are controlled at the desired levels. 

In a single stage design the error rates are designed to meet the desired levels at sample size $N$, the end of the study. In a group sequential setting the situation is more complex as we have (at least) two mutually dependent sets of errors to consider. Following convention we denote the probability of a type I error as $\alpha$, and similarly for type II as $\beta$, so we have error probabilities of $\alpha_1$ and $\beta_1$ at the interim look -- which may also be the final look depending on the results -- and then $\alpha_2$ and $\beta_2$ at the final look (end of the trial).  

The individual values of $\alpha_k$ and $\beta_k$ and total values, $\alpha=\sum^K_i{\alpha_i}$ and $\beta=\sum^K_i{ \beta_i}$, over the trial will likely need to meet specific design characteristics. For example, it's typically essential to ensure that $\alpha<0.025$ for a one-sided primary hypothesis in a pivotal trial. Note that here $\alpha$ now comprises of two sub components, $\alpha_1$ and $\alpha_2$ (two separate events) because it's possible to make a type I error at each look at the data. The error control can be on the total error as well as on the individual error components. If more than one interim look is desired then the number of $\alpha_i$ and $\beta_i$ terms increase in the obvious way.   
     
A group sequential design with one interim look has error rates that can be formalised as:
\begin{fleqn}
\begin{align}
&&\alpha_1&=P(Z_{1}>c_ {12}\mid{H_ 0});\label{eqne1}\\[2pt]
&&\alpha_2&=P(Z_{1}<c_ {12} \cap Z_{2}>c_ {2}\mid{H_ 0});\label{eqne2}\\[2pt]
&&\beta_1&=P(Z_{1}<c_ {11}\mid{H_ 1});\label{eqne3}\\[2pt]
&\text{and}&\beta_2&=P(Z_ {1}\in(c_{11},c_{12}) \cap Z_{2}<c_ {2}\mid{H_ 1});\label{eqne4}
\end{align}
\end{fleqn}
where $c_{12},c_{11},c_2$ are the critical values for stopping for efficacy (success) at interim, futility (failure) at interim and then at end of the study (where we have either success of failure) respectively. Given our use case hypothesis statement then $H_0$ is rejected for large positive values of $Z_i$, and $c_{12}>c_{11}$ as these are the efficacy and futility stopping boundaries respectively. The errors at the end of the study, $\alpha_2$ and $\beta_2$, are each joint events because in order to commit an error at the final look the trial cannot have already stopped at the interim look. As $Z_2$ is an accumulation of data which is already included in $Z_1$ then events at the end of the study are not independent from events at interim. The computation, therefore, of $\alpha_2$ and $\beta_2$ requires the joint probability distribution defined in eqns (\ref{eqj0})-(\ref{eqj2}). 

In group sequential designs the futility decision is often designated as \emph{non-binding}, which is the case in the events defined above. This can be seen from the $\alpha_2$ definition in eqn (\ref{eqne2}) which says that a type I error is possible at the end of the study only provided $Z_1<c_ {12}$ at interim. In other words, the type I error definition ignores the futility boundary at interim. A binding futility rule would change eqn (\ref{eqne2}) to $\alpha_2=P(Z_ {1}\in(c_{11},c_{12}) \cap Z_{2}>c_ {2}\mid{H_ 0})$ which now assumes that if $Z_1$ is below the futility boundary then the study will stop, if the study continues in this situation then the overall type I error will be inflated. Non-binding futility is arguably the more usual assumption in group sequential designs.  Binding or non-binding futility does not impact the type II error events. 

\subsubsection{Error spending}
We mentioned above that error control can be on the total error as well as on the individual error components. If the total error type I is first fixed, e.g. at $\alpha=0.025$, then in the terminology of a group sequential design we can decide where to ``spend'' this 0.025 worth of error across our multiple looks at the data. Similarly, for spending the type II error. 

Choosing how much error to spend at the interim look is a crucial part of creating an optimal group sequential design, as this determines the probability that the study might stop for efficacy at interim. Spending a small proportion of the total type I error at interim means that the study is less likely to stop for efficacy, but if it does stop the evidence is likely to be more persuasive. The converse is also true.  

There are many different approaches available in the literature in terms of how to distribute type I and type II errors within a group sequential design (see p.145 in Jennison\citep{Jennison:2000aa} and key articles therein\citep{Lan1983, Kim1987a,Hwang1990, DeMets:1994aa}).  Strictly speaking, ``error spending'' refers to a specific set of approaches for distributing the errors, whilst other classical approaches, such as that due to Wang-Tstais\citep{Wang1987}, similarly distribute error across interim looks but are methodologically distinct, and this approach includes as two special cases arguably the two most commonly referenced boundaries, the conservative O'Brien \& Fleming and less conservative Pocock boundaries. In our use case we consider the error spending approach of Kim\citep{Kim1987a} which is generalized in Jennison\citep{Jennison:2000aa} p. 148, and is one of those commonly used in practice. 

By construction, error spending approaches readily allow for additional unplanned interim looks to be included into a trial whilst still maintaining control of the overall error rates. This flexibility might not be needed but could be useful and such approaches perform as well as other ways of distributing errors. To see how error spending works, the expressions below use error spending from a family indexed by a parameter $\rho$ (see \citep{Jennison:2000aa} p.148). We have  
\begin{fleqn}
\begin{align}
&&\alpha_1&=f_\alpha(I_ {1}/I_{2}),&\label{spend1}\\
&&\alpha_2&=\alpha - f_\alpha(I_ {1}/I_{2})&\text{where }f_\alpha(t)=t^{\rho^\prime} \alpha,\label{spend2}\\
&&\beta_1&=f_\beta(I_ {1}/I_{2}),&\label{spend3}\\
&\text{and}&\beta_2&=\beta - f_\beta(I_ {1}/I_{2})&\text{where }f_\beta(t)=t^{\rho^{\prime\prime}}\beta.\label{spend4}
\end{align}
\end{fleqn}   
The two parameters $\rho^\prime$ and $\rho^{\prime\prime}$ determine respectively how much of the total type I and total type II errors are distributed between the interim and final look. If $\rho^\prime=1$ then the efficacy boundary will have similar properties to Pocock and if  $\rho^{\prime\prime}=3$ the futility boundary will have properties similar to O'Brien \& Fleming. The $I_i$ parameters are as previously, the information rates at each look.  

As can be seen from eqns (\ref{spend1})-(\ref{spend4}), adding the individual errors at interim and at the end of the study results in the cancellations of terms, and so ensures that the total error meets the design requirements. The spending functions -- $f_\alpha(t)$ and $f_\beta(t)$ -- use as the parameter $t$ the ratio of information at each interim look to the maximum information (i.e. at the end of the trial). This approach means that adding in a new unplanned interim look is straightforward in terms of controlling the error spend and maintaining the overall error target. 

Putting this all altogether for our use case, i.e. matching the design error requirements with the error spending scheme, then gives the following system of equations which require to be solved:
 
 \begin{fleqn}
\begin{align}
&&P(Z_{1}>c_ {12}\mid{H_ 0})&=(I_ {1}/I_{2})^{\rho^\prime} \alpha;\label{eqne1b}\\[2pt]
&&P(Z_{1}<c_ {12} \cap Z_{2}>c_ {2}\mid{H_ 0})&=\alpha -(I_ {1}/I_{2})^{\rho^\prime}\alpha;\label{eqne2b}\\[2pt]
&&P(Z_{1}<c_ {11}\mid{H_ 1})&=(I_ {1}/I_{2})^{\rho^{\prime\prime}} \beta;\label{eqne3b}\\[2pt]
&\text{and}&P(Z_ {1}\in(c_{11},c_{12}) \cap Z_{2}<c_ {2}\mid{H_ 1})&=\beta -(I_ {1}/I_{2})^{\rho^{\prime\prime}}\beta.\label{eqne4b}
\end{align}
\end{fleqn}

The system of equations (\ref{eqne1b})-(\ref{eqne4b}) encapsulates all the study design and sampling model parameters for our use case:
\begin{itemize}
 \item efficacy parameters $p_t, p_c$ and the non-inferiority margin $\delta$
 \item location of the interim analysis defined as a fraction, $\psi$, of overall per arm sample size $N$, also itself a parameter
 \item total type I and total type II errors $\alpha$ and $\beta$ respectively
 \item error spending power parameters, $\rho^\prime$ for type I error spending and $\rho^{\prime\prime}$ for type II error spending
 \item critical values for the test statistic stopping boundaries, $c_{11}, c_{12}, c_2$ 
 \end{itemize}
 giving a total of 12 parameters. We have four equations in 12 unknowns and so eight of these require to be fixed apriori in order to solve for the remaining unknowns.  Equations (\ref{eqne1b})-(\ref{eqne4b}) can be solved to determine the $N$ required to meet given target errors or similarly solve for power given a fixed sample size (so swap unknown $N$ for unknown $\beta$). In the results section we show an example of how to solve this system of equations in order to compute sample size and power.

\subsubsection{Crossing probabilities}\label{stopping}
One of the additional parameters or operating characteristic of a design which has an interim look is the probability of stopping the study for efficacy (success) at that look. What constitutes a reasonable lower bound on this probability is case specific, for example if there is considerable uncertainty in the treatment effect then the interim may be to capture this information and so there is no target value for this parameter. Alternatively, it may be that our prior belief in regard to likely efficacy is such that planning a design where the probability of stopping is an explicit target design characteristic may be appropriate. This operating characteristic is also referred to as a crossing probability and is defined in the obvious way from the above events:
 \begin{fleqn}
\begin{align}
&&\omega_1&=P(Z_{1}>c_{12}\mid{H_1})\label{c1}\\
&\text{and}&\omega_2&=P(Z_{1}\in(c_{11},c_{12}) \cap Z_{2}>c_{2}\mid{H_1})\label{c2}
\end{align}
\end{fleqn}
where eqn (\ref{c1}) is the probability of stopping at interim for efficacy, i.e.~the probability of crossing into the null hypothesis rejection region under the alternative hypothesis, and eqn (\ref{c2}) is the probability of not stopping at interim (for efficacy or futility) and then achieving success for efficacy at the final look. The sum of these two events equals the power of the design, i.e.~overall probability of success whether this happens at interim or at the end of the study. Similar crossing probabilities can be computed for early stopping due to futility.

\subsubsection{Minimum difference}
Introducing an interim analysis to allow for early stopping can be a pragmatic way to de-risk a trial, however, it must also provide persuasive evidence that the treatment effect is robust, given that the trial is stopping (for success) earlier than planned. The FDA guidance on adaptive designs \citep{FDA_guidance} states ``stopping rules require highly persuasive results in terms of both the magnitude of the estimated treatment effect and the strength of evidence of an effect''. One interpretation of this text is that it is advising against stopping for statistical significance on its own, and what is additionally required is that the treatment magnitude should also be clinically persuasive. 

One approach for quantifying treatment magnitude at an interim look in a group sequential study is to ask: what is the minimum observed treatment effect that would cause the test statistic to fall into the type I error rejection region at the interim look? We refer to this as the minimum difference and this can be computed at both an interim look and final look. 

From the expression defining the test statistic (eqn \ref{eqn1}) we have
\begin{fleqn}
\begin{align}
&&d_1&=\hat{p}_t-\hat{p}_c= \delta+ c_{12} \sqrt{I_{1}}^{-1}\label{eqmin1}\\
&\text{and}&d_2&=\hat{p}_t-\hat{p}_c= \delta+ c_2 \sqrt{I_{2}}^{-1}.\label{eqmin2}
\end{align}
\end{fleqn}
where $\hat{p}$ denotes the observed sample estimate. These expressions show that the more conservative the error spending rule (so larger critical values) then the larger the minimum difference all else being equal. We can also see that as the sample size becomes large then the minimum difference converges to the non-inferiority margin as it should. 

\subsection{Optimizing operating characteristics}\label{optmeth}
In the previous sections we have shown how to compute the typical operating characteristics needed for a study, such as sample size and power, given known -- fixed or assumed -- values for all other design parameters, excluding critical values. We can go further and allow some of these previously fixed parameters to be unknown, which also then allows us to target additional operating characteristics. This then brings us into the area of multi-aspect design planning optimization where we wish to find a design which simultaneously meets multiple target operating characteristics beyond power and type I error.  

Recall that the general numerical problem to be solved in group sequential planning scenarios is that we have a system of equations and a set of unknowns. Provided the problem is not over-parameterized then we can mix and match which parameters are fixed and which are to be estimated. This can be thought of as numerical optimization, as we are letting unknown parameters be estimated via an algorithm in order that other target design characteristics are met. Conceptually, this is simply an extension of the analogous single stage study design situation where we fix $N$ and estimate power or vice-versa, whilst maintaining control of the type I error. In a group sequential design, however, we have more parameters and design characteristics along with multiple dependent test statistics so is numerically more difficult than single stage designs.    

A practically important example in a group sequential setting is where as well as meeting a given overall minimum power, the design should also achieve a minimum target for the crossing probability for stopping for efficacy at interim, and additionally, also achieve a minimum difference target. To design a study with these three target design characteristics (including additionally a maximum overall type I error, which is implicit as a given essential requirement) then we have the following systems of equations in (\ref{eqne1bb})-(\ref{eqmin1b}), 
 \begin{fleqn}
\begin{align}
&&P(Z_{1}>c_ {12}\mid{H_ 0})&=(I_ {1}/I_{2})^{\rho^\prime} \alpha,\label{eqne1bb}\\[2pt]
&&P(Z_{1}<c_ {12} \cap Z_{2}>c_ {2}\mid{H_ 0})&=\alpha -(I_ {1}/I_{2})^{\rho^\prime}\alpha,\label{eqne2bb}\\[2pt]
&&P(Z_{1}<c_ {11}\mid{H_ 1})&=(I_ {1}/I_{2})^{\rho^{\prime\prime}} \beta,\label{eqne3bb}\\[2pt]
&&P(Z_ {1}\in(c_{11},c_{12}) \cap Z_{2}<c_ {2}\mid{H_ 1})&=\beta -(I_ {1}/I_{2})^{\rho^{\prime\prime}}\beta,\label{eqne4bb}\\
&&\omega_1&=P(Z_{1}>c_{12}\mid{H_1}),\label{c1b}\\
&\text{and}&d_1&=\hat{p}_t-\hat{p}_c= \delta+ c_{12} \sqrt{I_{1}}^{-1}.\label{eqmin1b}
\end{align}
\end{fleqn}
These equations are similar to those in (\ref{eqne1b})-(\ref{eqne4b}) but we now have two additional equations, one for the crossing probability and one for the minimum difference.

To achieve these target operating characteristics we require to solve the system of equations in (\ref{eqne1bb})-(\ref{eqmin1b}). To do this we need to free up some moving parts (let previously fixed parameters be unknown and free to be estimated), and two candidates for this are the error spending scheme, e.g. for efficacy, $\rho^\prime$, and $\psi$, the fraction of the overall sample size where we conduct the interim look. This then gives us six equations in six unknowns. We show an example of how this may be achieved and the associated results and outputs in the results section.

\subsection{Numerical Method}
In order to solve the system of non-linear equations in (\ref{eqne1bb})-(\ref{eqmin1b}) or similar more complex examples, such as with additional interim looks, we need a multi-dimensional root finding algorithm. Such approaches are available in open source libraries and platforms and typically implement variations of Newton's method (e.g. see Mor\'e\citep{More1979}). Multi-dimensional root finding algorithms are only locally convergent when applied to non-linear systems, which means that initial starting values for a numerical search can be important to avoid convergence issues. An additional complexity in the non-linear systems in group sequential methods is that they also contain embedded multi-dimensional integrals, and whose integration limits are parameters in the root-finding algorithm which can lead to numerical stability problems. 

Our case study results in the following section use a hybrid numerical approach where root-finding was applied to a simpler system than in (\ref{eqne1bb})-(\ref{eqmin1b}), and where some parameters were not included in the root-finding, but were rather fixed across a grid. At each point in this grid the root-finding algorithm was run to solve for the remaining unknown parameters. This provided more numerically stable results than directly trying to solve (\ref{eqne1bb})-(\ref{eqmin1b}), albeit at some additional computational cost. Full details and R code are provided in the supplementary material.

\section{Results}\label{casestudy}
In this section we provide numerical examples of how to calculate design operating characteristics from first principles for a group sequential design with a single interim look, and with the primary hypothesis test and test statistics as detailed above. All R code necessary for generating the results presented here is provided in the supplementary material, with a code snippet provided later in this section showing key parts. 

\subsection{Visualising a group sequential design}
\begin{figure}[htbp]
	\centering
	\includegraphics[width=\linewidth]{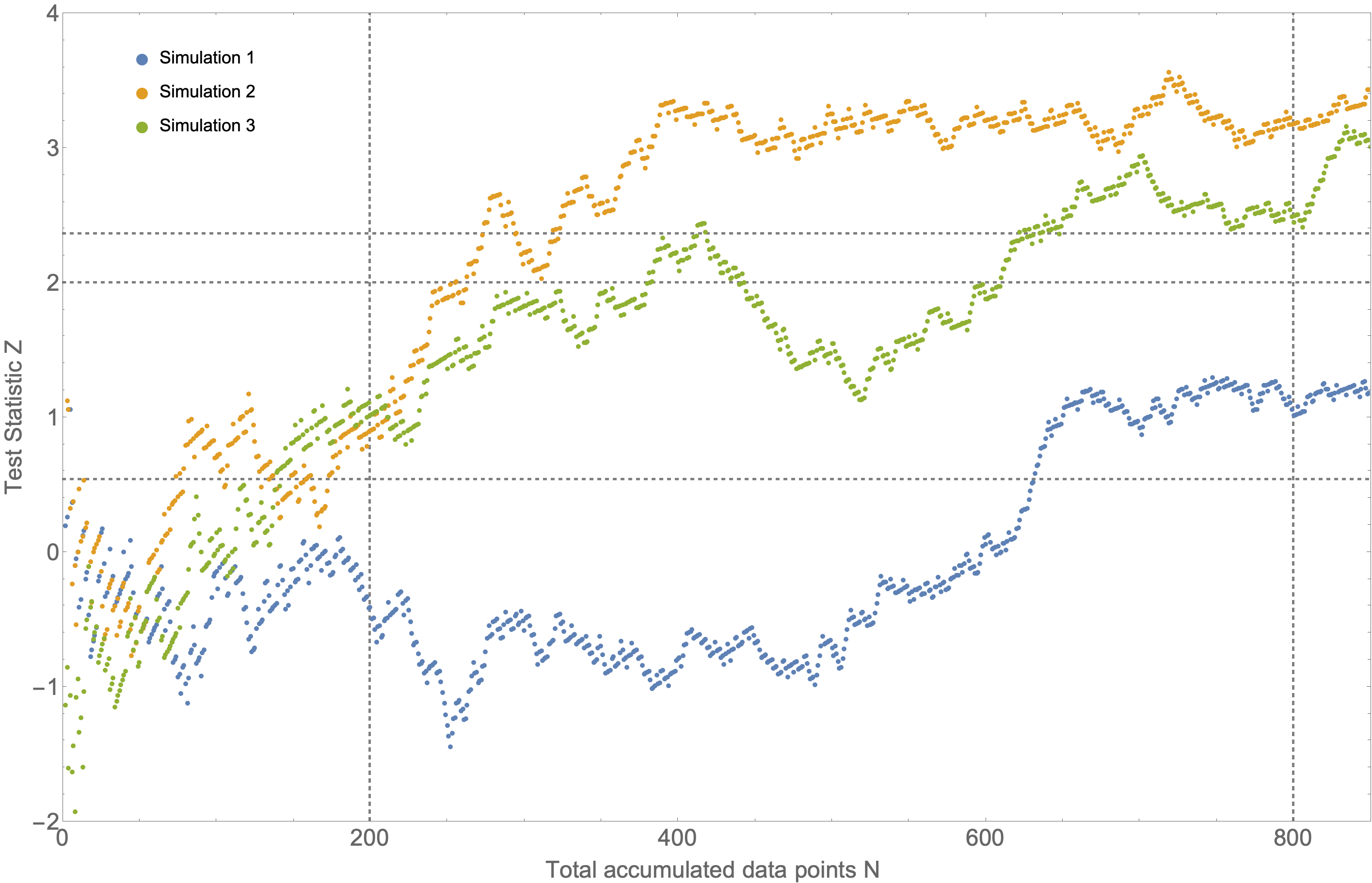}
	\caption{Three realisations of the test statistic $Z_i$ in eqn (\ref{eqn1}) as the trial accumulates new data (patients), see main text for the parameters used.  The vertical lines indicate where looks at the data may occur, where these can be moved left or right depending on how much data is desired to be accumulated before any analyses, the first for an interim look and the second for the position of the final analysis. Currently shown is an interim analysis at $n=200$ (per arm) and a final analysis at $n=800$. The three horizontal lines indicate critical value cut-offs for the test statistic which can be moved up or down depending on the operating characteristics required.} 
	\label{fig1}
\end{figure}

Before we present numerical results of sample size and power calculations it is first instructive to visualize the behaviour of the $Z_i$ test statistics over $i$. Figure \ref{fig1}
shows three simulations of how the random variable $Z_i$ might behave as a trial accumulates each new data point under the alternative hypothesis\footnote{technically each pair of data points, as here for simplicity $i$ is the accumulation of a pair of patients, one for each treatment arm, and we are assuming fixed 50/50 recruitment balance}, where $i=1,\dots,800$, for each of the three realizations of $Z_i$ shown (and in a slight abuse of notation, here $i$ denotes the accumulated sample size itself, rather than indexing interim looks as in eqn \ref{eqn1}). Each of these simulations uses the same parameter set $(p_t=0.58, p_c=0.6, \delta=-0.1)$ but a different random seed and so the differences in each trajectory are only due to stochasticity. Under the alternative hypothesis each trajectory will on average drift upwards as data accumulates. The vertical lines show where an interim and final analysis may take place, and the horizontal lines represent critical value cut-off points, where the upper of these would be for the interim look. Using the critical values as shown in the figure then we can see that simulation one is below the futility bound at the interim, and such a low observed value could cause a study to stop for futility. In contrast, simulations two and three are in the continuation zone at interim and both also exceed the efficacy (success) boundary at the end the study, whereas simulation one is a type II error.   

\subsection{R code snippet}
Listing 1 contains indicative R code demonstrating how to calculate the various type I and type II errors in a group sequential design, along with the crossing probability and minimum difference. Also shown is an example of multi-dimensional root finding in R. Complete R code is provided in the supplementary material.   

\begin{lstlisting}[language=C,float,caption={R code snippet for error spending and root finding. Parameter names have been chosen to match those in the main text.}]
library(mnormt)
library(rootSolve)

## compute Type I and Type II errors

## type I at interim
A1<-sadmvn(lower=c(c12,-Inf), upper=c(Inf,Inf),  mean=meanH0(...), 
                                                 varcov=covz1z2(...))
## type I at end of study 
A2<-sadmvn(lower=c(-Inf,c2), upper=c(c12,Inf),   mean=meanH0(...), 
                                                 varcov=covz1z2(...))
## type II at interim
B1<-sadmvn(lower=c(-Inf,-Inf), upper=c(c11,Inf), mean=meanH1(...), 
                                                 varcov=covz1z2(...))
## type II at end of study 
B2<-sadmvn(lower=c(c11,-Inf), upper=c(c12,c2),   mean=meanH1(...), 
                                                 varcov=covz1z2(...))
## crossing probability
CR1<-sadmvn(lower=c(c12,-Inf), upper=c(Inf,Inf), mean=meanH1(...), 
                                                 varcov=covz1z2(...))
## min diff
DF1<- d+c12/(sqrt(I_k())) # d=NI margin, I_k() computes the information

## type I spend at interim
A1_S<- ((I_k(p_t,p_c,n_z1)/I_k(p_t,p_c,n_z2))^rho1)*alpha

## type I spend at end of study  
A2_S<- alpha-((I_k(p_t,p_c,n_z1)/I_k(p_t,p_c,n_z2))^rho1)*alpha

## type II spend at interim  
B1_S<- ((I_k(p_t,p_c,n_z1)/I_k(p_t,p_c,n_z2))^rho2)*beta

## type II spend at end of study  
B2_S<- beta-((I_k(p_t,p_c,n_z1)/I_k(p_t,p_c,n_z2))^rho2)*beta

## for root finding 
result<-c(f1=A1-A1_S,f2=A2-A2_S,f3=B1-B1_S,f4=B2-B2_S)

## call to find roots - find c11, c12, c2, N
multiroot(f=myResultFunction,
          start=c(c11_init,c12_init,c2_init,N_init),
          parms=myExtraPars)

\end{lstlisting}

\subsection{Determining sample size and power}
To calculate sample size for our use case we implement (\ref{eqne1b})-(\ref{eqne4b}) into R code and apply the following parameter values: $(p_t=0.58, p_c=0.6, \delta=-0.1,n_1=\psi N, \psi=0.6,\rho^{\prime}=2, \rho^{\prime\prime}=3)$, where $n_1$ is the per arm sample size at the interim look and $N$ the total per arm sample size. In addition we set $\alpha=0.025$ and $1-\beta=0.9$. Using similar code to listing 1 we then get as results $(c_{11}=0.548,c_{12}=2.366,c_2=2.04,N=831.6$), which gives a crossing probability for efficacy at interim of 0.58 and a minimum difference of -0.026, and error spend (p-value boundary at interim) for efficacy of 0.009. These results were checked against Cytel's EAST. 

To compute power we use near identical code, all that is required is switching the parameter $N$ to fixed and $\beta$ as a parameter to be estimated via root finding. Fixing $N=831.6$ and computing power then gets back the same results as above.

\subsection{Multi-aspect examples - power, crossing probability and minimum difference}
\begin{figure}[htb]
	\centering\caption{Multi-aspect example 1 - fixed power at 90\%, (a) crossing probability for success at interim, (b) minimum difference at interim, (c) total sample size per arm, (d) p-value boundary at interim. For details of highlighted points see main text.} 
	\vspace*{0.3cm}
	\includegraphics[width=\linewidth]{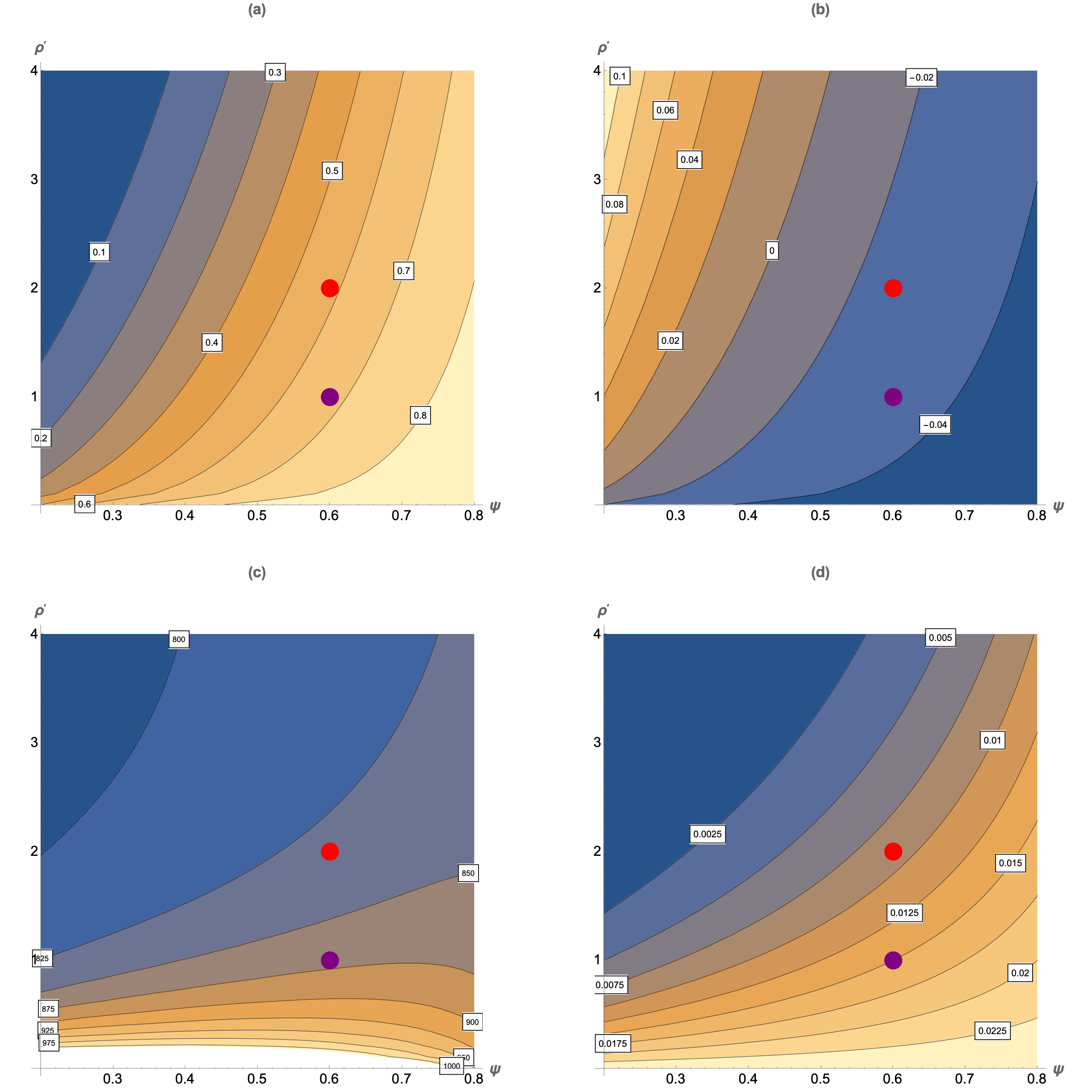}
	
	\label{fig2}
\end{figure}

\begin{figure}[htb]
	\centering\caption{Multi-aspect example 2 - fixed N=831.6, (a) crossing probability for success at interim, (b) minimum difference at interim, (c) total sample size per arm, (d) p-value boundary at interim. For details of highlighted points see main text} \label{fig3}
	\vspace*{0.3cm}
	\includegraphics[width=\linewidth]{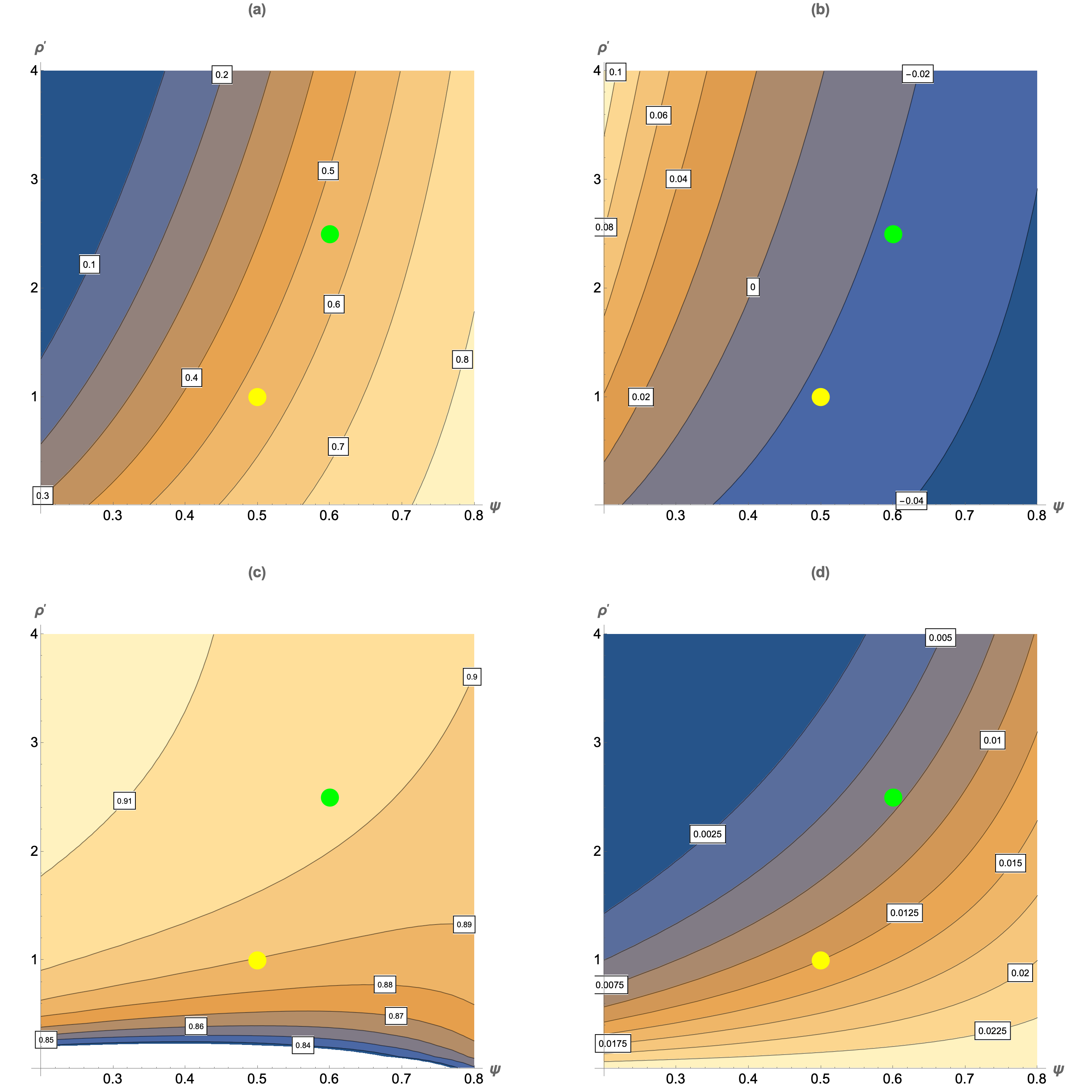}
\end{figure}

In Section \ref{optmeth} we detailed the expressions needed for designing a study to meet multiple simultaneous target operating characteristics, specifically power, success at interim look (crossing probability for efficacy), minimum  difference and type I error. In figures \ref{fig2} and \ref{fig3} we show root-finding results across a 2-D surface where the location of the interim look (parameter $\psi$, the fraction of total $N$ at interim) and error spending at interim on efficacy (parameter $\rho^{\prime}$) vary.  

In figure \ref{fig2} 
the power is fixed at $1-\beta=0.9$, with type II error spending fixed at $\rho^{\prime\prime}=3$, and overall type I $\alpha=0.025$. The efficacy parameters are as in figure \ref{fig1}, $(p_t=0.58, p_c=0.6, \delta=-0.1)$. Each of the four panels shows how different operating characteristics vary with $\psi$ and $\rho^{\prime}$ subject to our parameter constraints. Two points are highlighted on this figure, one with $(\psi,\rho^{\prime})=(0.6,2)$ and a second with $(\psi,\rho^{\prime})=(0.6,1)$, where the latter is a more aggressive error spending rule (similar to Pocock). From panels (a)-(d) in figure \ref{fig2} we can see that at the more aggressive combination of $(\psi,\rho^{\prime})$: the crossing probability increases from 0.58 to 0.68; the minimum difference decreases from -0.026 to -0.034; the total per arm sample size increases from 831.6 to 869.4; and the type I error spend at interim increases from 0.009 to 0.015. 

Figure \ref{fig3}
 is similar to \ref{fig2} but now the design fixes the total per arm sample size at 831.6 (the value corresponding to 90\% power in Figure \ref{fig2} with $\psi=0.6$ and $\rho^{\prime}=2$) rather than fixing the power. Two points are highlighted on this figure, one where $(\psi,\rho^{\prime})=(0.6,2.5)$ and another where $(\psi,\rho^{\prime})=(0.5,1)$. The latter is a more aggressive error spending rule but also brings forward the interim to 50\% of recruitment. By again comparing the various panels in the plot we can see that the more aggressive spending rule coupled with earlier recruitment largely cancel each other out in terms of overall impact on the operating characteristics: the crossing probability reduces slightly from 0.554 to 0.542; the minimum difference stays (approximately) the same at -0.023; the overall power decreases slightly from 0.903 to 0.89; and the type I error spend (p-value boundary for efficacy) at interim increases from 0.007 to 0.0125.   

Figures \ref{fig2} and \ref{fig3} are examples of multi-aspect design planning - we attempt to simultaneously optimize multiple design parameters by varying others in order to achieve an optimal trade-off of each. This is analogous to typical study planning with a single stage design where we trade-off sample size against power while keeping the type I error fixed. Planning is more complex in group sequential designs as they have more parameters and more operating characteristics which we may wish to ensure achieve at least minimum bounds in any final design. Our current examples include only a single interim look, additional interim looks can be analysed in a similar way.  

The two examples shown here vary the location of the interim look ($\psi$) and the type I error spending ($\rho^{\prime}$). This is just one possibility of many, other options could include varying the type II error spend, and/or including the crossing probability for futility as an explicit operating characteristic, in addition to varying the efficacy parameters. Which parameters to keep fixed, either apriori (such as the assumed efficacy parameters which we kept fixed here) or which to vary across a grid, or determine via root-finding is problem specific. Although some parameters more naturally fall into one category than the other, for example it seems appropriate to determine critical values via root-finding rather than across a grid, as we would not typically directly choose critical values, rather these are set as a byproduct of other design requirements.

\section{Discussion}
We have provided a short tutorial in group sequential methods, showing how some of the key operating characteristics are calculated using case study examples. The specific situation considered was a trial with a non-inferiority primary hypothesis and a binary endpoint, but the concepts and general approach are widely applicable to other types of objectives and endpoints. See Jennison and Turnbull\citep{Jennison:2000aa} for a selection of other application areas.
  
From a statistical perspective there are (at least) two key areas of focus from a planning point of view when introducing an interim look into a study design. Firstly, robustness of any interim analysis. If an interim look is planned to be conducted early in a trial then the sampling distributions of the test statistics may be poorly approximated by the standard analytical formulation (joint normal density with covariances as ratios of information rates). How to determine this in practice can be difficult, with simulation being the obvious option. However, for pivotal studies at least, generally speaking sample sizes are likely to be more than sufficient for good asymptotic approximations to hold, e.g. hundreds of patients, without requiring simulations to verify this. But exceptions will exist. 

A second key area of focus is stakeholder engagement, education and alignment in interpreting and agreeing the various desired operating characteristics in a group sequential design. This is arguably the most challenging part of designing a study. An interim look might be included in a trial design because their is a strong aspiration to stop the study early for success, and given uncertainty around the apriori estimated treatment effect, then a group sequential design could be a good option. However, it is then necessary to partner with key stakeholders, such as clinical leads, so that the interpretation of the group sequential operating characteristics at interim are clearly understood, and their personal interpretations of risk can be clearly translated into, e.g. target crossing probabilities and minimal differences. 

The use of specialist software, such as EAST or Solara by Cytel or PASS by NCSS, makes group sequential designs readily accessible to trial statisticians. Such packages may not give the full flexibility of using bespoke code, such as that shown in our case study examples, but still provide a wide range of features suitable for most use cases. Alternatively, one of the advantages of using bespoke code is that it's possible to mix and match parameters: we have a non-linear system of equations which can be solved for whatever combination of known and unknowns are desired. For example, earlier we solved to find a design which met simultaneously a crossing probability target and a minimal difference target. This full flexibility can be advantageous but also comes with the usual risks associated of using ad-hoc code, and the responsibility which comes with this in terms of quality assurance in a Good Clinical Practice (GCP) environment. That said, and as seen above, the actual calculations are not particularly onerous in platforms such as R. 

All results presented here can be reproduced using the code and information provided in the supplementary material and available on GitHub at <author or GSK repository https:// >.

\bibliographystyle{unsrtnat}
\bibliography{references}  






\end{document}